\pdfoutput=1

\documentclass[11pt,twoside,a4paper,cmspaper,final,collab]{cms-tdr}
\def\svnVersion{79257:79406}\def\cmsCernNo{2011-139}\def\cmsCernDate{\today}\def\cmsMessage{Submitted to Physical Review Letters}

\begin{document}\cmsNoteHeader{EXO-11-005}

\hyphenation{had-ron-i-za-tion}
\hyphenation{cal-or-i-me-ter}
\hyphenation{de-vices}

\hyphenation{env-iron-men-tal}%    just an example
\hyphenation{had-ron-i-za-tion}
\hyphenation{cal-or-i-me-ter}
\hyphenation{de-vices}
\RCS$Revision: 79406 $
\RCS$HeadURL: svn+ssh://alverson@svn.cern.ch/reps/tdr2/papers/EXO-11-005/trunk/EXO-11-005.tex $
\RCS$Id: EXO-11-005.tex 79406 2011-09-23 01:38:13Z alverson $
\newcommand{\bigT}{\ensuremath{\cmsSymbolFace{T}}\xspace}
\newcommand{\bigaT}{\ensuremath{\overline{\cmsSymbolFace{T}}}\xspace}
\newcommand{\TTbar}{\bigT\bigaT}
\newlength\cmsFigWidth
\ifthenelse{\boolean{cms@external}}{\setlength\cmsFigWidth{\columnwidth}}{\setlength\cmsFigWidth{0.6\textwidth}}
\cmsNoteHeader{11-005}
\title{Search for a Vector-like Quark with Charge $2/3$ in $\cPqt + \cPZ$ Events from $\Pp\Pp$ Collisions at $\sqrt{s} = 7$ TeV}

\author{The CMS Collaboration}

\date{\today}

\abstract{
A search for pair-produced heavy vector-like charge-2/3 quarks, \bigT, in $\Pp\Pp$
collisions at a center-of-mass energy
of $7\TeV$, is performed with the CMS detector at the LHC.
Events consistent with the flavor-changing-neutral-current decay of a \bigT quark to a top quark
and a $\cPZ$ boson are selected by requiring
two leptons from the $\cPZ$-boson decay, as well as an additional isolated charged lepton.
In a data sample corresponding to an integrated luminosity of $1.14\fbinv$,
the number of observed events is found to be consistent with the standard model background prediction.
Assuming a branching
fraction of 100\% for the decay $\bigT \to \cPqt\cPZ$, a \bigT quark with a mass less than
475\GeVcc is excluded at the 95\% confidence level.
}

\hypersetup{%
pdfauthor={CMS Collaboration},%
pdftitle={Search for a Vector-like Quark with Charge 2/3 in t + Z Events from pp Collisions at sqrt(s) = 7 TeV},%
pdfsubject={CMS},%
pdfkeywords={CMS, physics}}

\maketitle %maketitle comes after all the front information has been supplied

Recently, there has been renewed interest in the search for fourth-generation particles~\cite{Holdom:2009rf} that could have
escaped the stringent bounds set by precision measurements~\cite{tagkey2006257,Nakamura:2010zzi}.
Searches for
$\cPqb^\prime \to \cPqt\PW$~\cite{Aaltonen:2011vr,CMS:bprime2tW} and $\cPqt^\prime \to \cPqb\PW, \cPq\PW$~\cite{Abazov:2011vy} decays
have been performed at the Tevatron and LHC,
setting lower bounds on the masses of fourth-generation quarks $\cPqb^\prime$ and $\cPqt^\prime$.
The decays $\cPqb^\prime\to \cPqb\cPZ$ and $\cPqt^\prime\to \cPqt\cPZ$ are flavor-changing-neutral-current (FCNC) processes and,
since they proceed through loop diagrams, they are expected~\cite{Arhrib:2006pm} to have branching fractions of
$\mathcal{O}(10^{-5}$--$10^{-4})$.
Lower bounds on the mass of a $\cPqb^\prime$ decaying to $\cPqb\cPZ$ have been established~\cite{CDF:bprime2bW}.
If a vector-like quark of charge~2/3 (denoted~\bigT) exists, however, as
expected in several models of new physics~\cite{TopLikeMode-compositeHiggs,TopLikeMode-ExtraDimensions,TopLikeMode-LittleHiggs},
it would have tree-level FCNC couplings that could result in a large branching fraction for FCNC \bigT decays.
For example, for a vector-like \bigT with a new Yukawa coupling~\cite{BR-tree_level,TopPartners},
the decays $\bigT \to \cPqt\cPZ$ and $\bigT \to \cPqt\PH$ could be dominant, where H is the Higgs boson.
If the Higgs decay channel is kinematically forbidden, the  $\bigT \to \cPqt\cPZ$
branching fraction could be close to 100\%.

In this Letter, we report the results of a first search for pair-produced \bigT quarks
that decay to top quarks and $\cPZ$ bosons, with the Compact Muon Solenoid (CMS) detector at the Large Hadron Collider (LHC).
The decay chain, $\Pp\Pp \to \TTbar X$, with $\TTbar \to \cPqt\cPZ\cPaqt\cPZ\to \cPqb\cPaqb\PWp\PWm\cPZ\cPZ$,
can generate a very clean signature if at least one $\cPZ$ boson decays to $\ell^+\ell^-$,
where $\ell$ is an electron or a muon,
and the decay of one of the $\PW$ bosons yields an additional isolated charged lepton.
A search for singly-produced vector-like quark has been performed by the D0 collaboration~\cite{singleVector-like}.

The central feature of the CMS apparatus is a superconducting
solenoid that provides an axial magnetic field of 3.8 T.
Charged
particle trajectories are measured within the field volume by pixel and silicon strip
tracker. The calorimeter enclosing the tracker includes a lead tungstate
crystal electromagnetic calorimeters (ECAL),
which is composed of a barrel part and two endcaps, a lead and silicon preshower detector in front of the ECAL endcaps,
and a brass/scintillator hadron calorimeter (HCAL) that together provide an
energy measurement for electrons, photons, and hadronic jets.  Muons are
identified and measured in gas-ionization detectors embedded in the
steel return yoke outside the solenoid. The detector is nearly
hermetic, allowing accurate energy balance measurements in the plane
transverse to the beam direction.  The direction of particles measured
inside the CMS detector is described using the azimuthal angle ($\phi$) and the
pseudorapidity ($\eta$), which is defined as $\eta\equiv-\ln[\tan\theta/2]$,
where $\theta$ is the polar angle relative to the counterclockwise
proton beam direction, as measured from the nominal interaction
vertex. A more detailed description of the CMS detector can be found
elsewhere~\cite{CMS}.

This study is based on a sample of pp collisions as $\sqrt{s}=7$\TeV recorded in March-June 2011,
and corresponds to an integrated luminosity of ($1.14\pm0.05\fbinv$).  The CMS
trigger system consists of hardware and software
triggers~\cite{HLT} that are used to select events for further
analysis.
Events selected for this search are required to pass one of several dilepton triggers.
The efficiencies of the dilepton triggers are measured using an independent data sample collected with a jet-based trigger
and containing at least two fully reconstructed leptons,
and found to be 99\% for two-electron, 89\% for two-muon, and 97\% for electron-muon triggers.

Muon candidates are required to have a
transverse momentum $\pt>15~\GeVc$
and be within the fiducial range
$|\eta|<2.4$. The reconstructed muon track must be associated with signals in the pixel
and silicon strip detectors, as well as track segments in the muon system,
and have a high-quality global fit using the information of both the central tracker and the muon detector.
The muon reconstruction is described in detail in Ref.~\cite{MUON}.
The muon candidate is also required to be consistent with
coming from the primary interaction vertex~\cite{IP}.

Electron candidates are reconstructed using clusters of energy
deposits in the ECAL that are matched to a track reconstructed in the
tracker.  A candidate is required to have $\pt>20\GeVc$ and
be within the fully instrumented barrel ($|\eta|<1.44$) or endcap
($1.57<|\eta|<2.5$) regions. The track must also be consistent with originating from the
interaction vertex.  Electrons are identified based on the ratio between the
energy depositions in the ECAL and the HCAL, the shower width in
$\eta$, and the distance between the energy-weighted mean position in
the ECAL and the extrapolated position of the associated track
measured in both $\eta$ and $\phi$.  The selection criteria are optimized
to identify electrons from $\PWm$~or $\cPZ$-boson decays with an efficiency of $85\%$,
while suppressing at least 98\% of
candidates originating from hadronic jets~\cite{EID}.

Leptons from $\PWm$~or $\cPZ$-boson decays tend to be
isolated from other particles in the event.
Several requirements are imposed on the sum of the transverse momentum or energy of
particles (not including the lepton itself) surrounding the lepton
within a cone of $\Delta R\equiv\sqrt{(\Delta\eta)^2 +
(\Delta\phi)^2}=0.3$,
where $\Delta\eta$ and $\Delta\phi$ are the differences in pseudorapidity and azimuthal angle between the lepton and
the particle directions.
The sum of the $\pt$ of tracks surrounding a muon
candidate must be less than $3\GeVc$.  Similarly, an electron
candidate in the barrel (endcaps) is rejected if the sum of the $\pt$ of
tracks around it is greater than $9\%$~($5\%$) of the electron's
$\pt$, the sum of the $\et$ in the surrounding ECAL region is greater than
$8\%$~($5\%$) of that of the candidate, or if the sum of the $\et$ in the
surrounding HCAL is greater than $10\%$~($2.5\%$) of the electron's
$\et$.
Electron candidates within a cone of $\Delta R=0.1$ of a muon
candidate are rejected in order to remove misidentified muon bremsstrahlung photons mistakenly associated
with the muon-candidate track and misidentified as electrons.
Electrons identified as resulting from photon conversions are also
rejected.

Jets are reconstructed from particles whose identities and
energies have been determined by a particle-flow
technique~\cite{PFT-09-001,PFT-10-001}.
All particles found by the particle-flow algorithm
are clustered into jets using the
anti-$k_{T}$ algorithm with the distance parameter
of 0.5~\cite{anti-kT}.
Jet energies are
corrected
for non-uniformity in calorimeter
response and for differences found between
jets in simulation and data~\cite{Chatrchyan:2011ds}.
Jet candidates are required to
have $\pt>25~\GeVc$, be within $|\eta|<2.4$, and pass quality
requirements that reject most misidentified jets arising from calorimeter noise.
Jets must also be separated from all lepton candidates by a
distance $\Delta R>0.4$.

We select events that contain at least one well-reconstructed
interaction vertex and
a leptonic $\cPZ$-boson decay, which is identified by
requiring oppositely charged, same-flavor leptons (e or $\mu$) having an invariant mass
in the range $60<M_{\ell^+\ell^-}<120~\GeVcc$.
At least three leptons and at least two jets are required.
An additional reduction of the standard model (SM) background is obtained by requiring %$R_{\rm T}>80~\GeVc$.

\begin{align}
R_{\bigT}\equiv\sum\limits_{i\neq1,2}\pt(\text{jet}_{i})+\sum\limits_{j\neq1,2}\pt(\text{lepton}_{j})>80\GeVc,
\end{align}
where the $i,\ j\neq1,2$ indicates that the sum extends over all leptons and jets, except the two highest-$\pt$ ones.

\begin{figure*}[t]
  \begin{center}

    \includegraphics[width=0.3\textwidth]{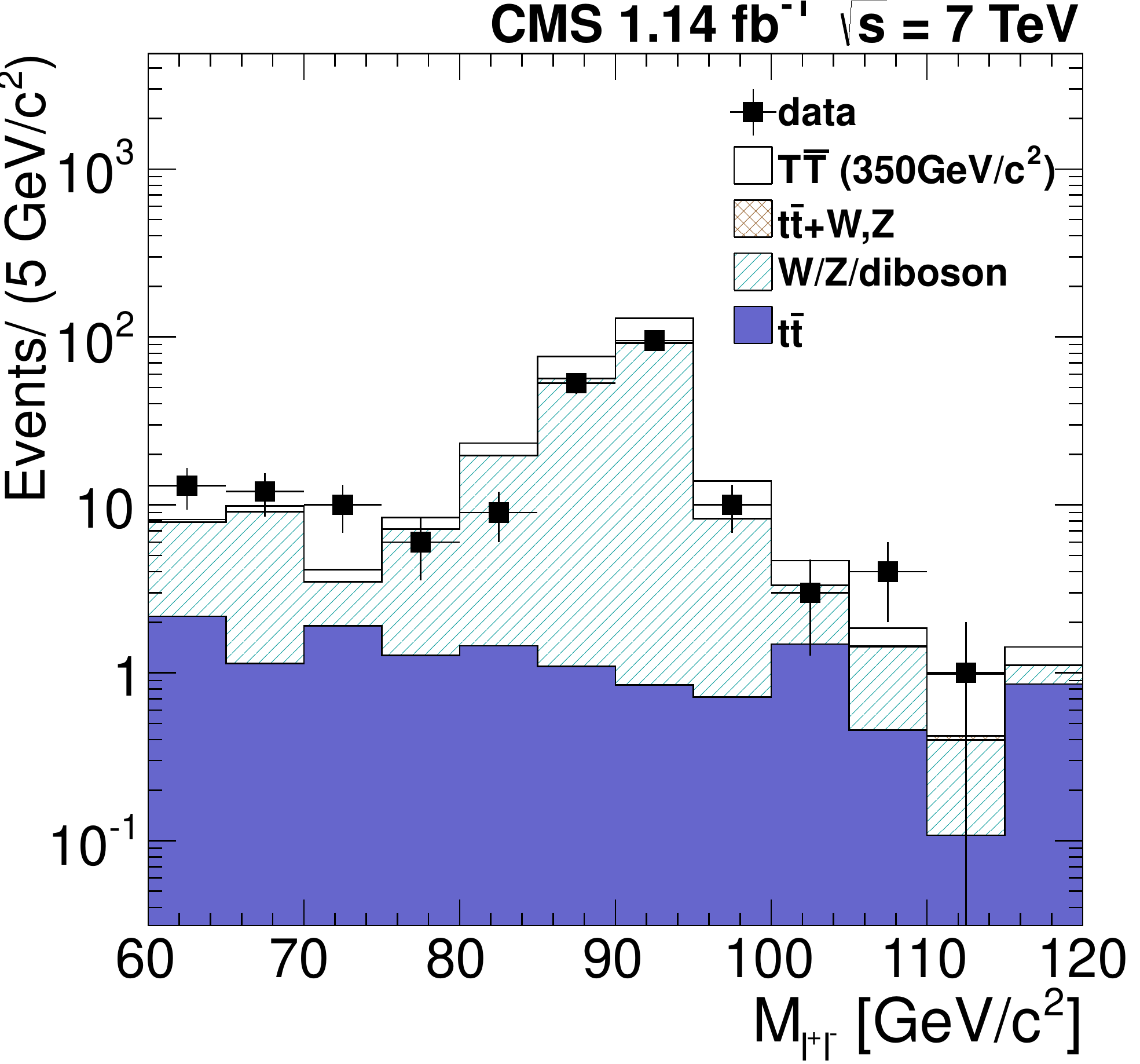}
    \includegraphics[width=0.3\textwidth]{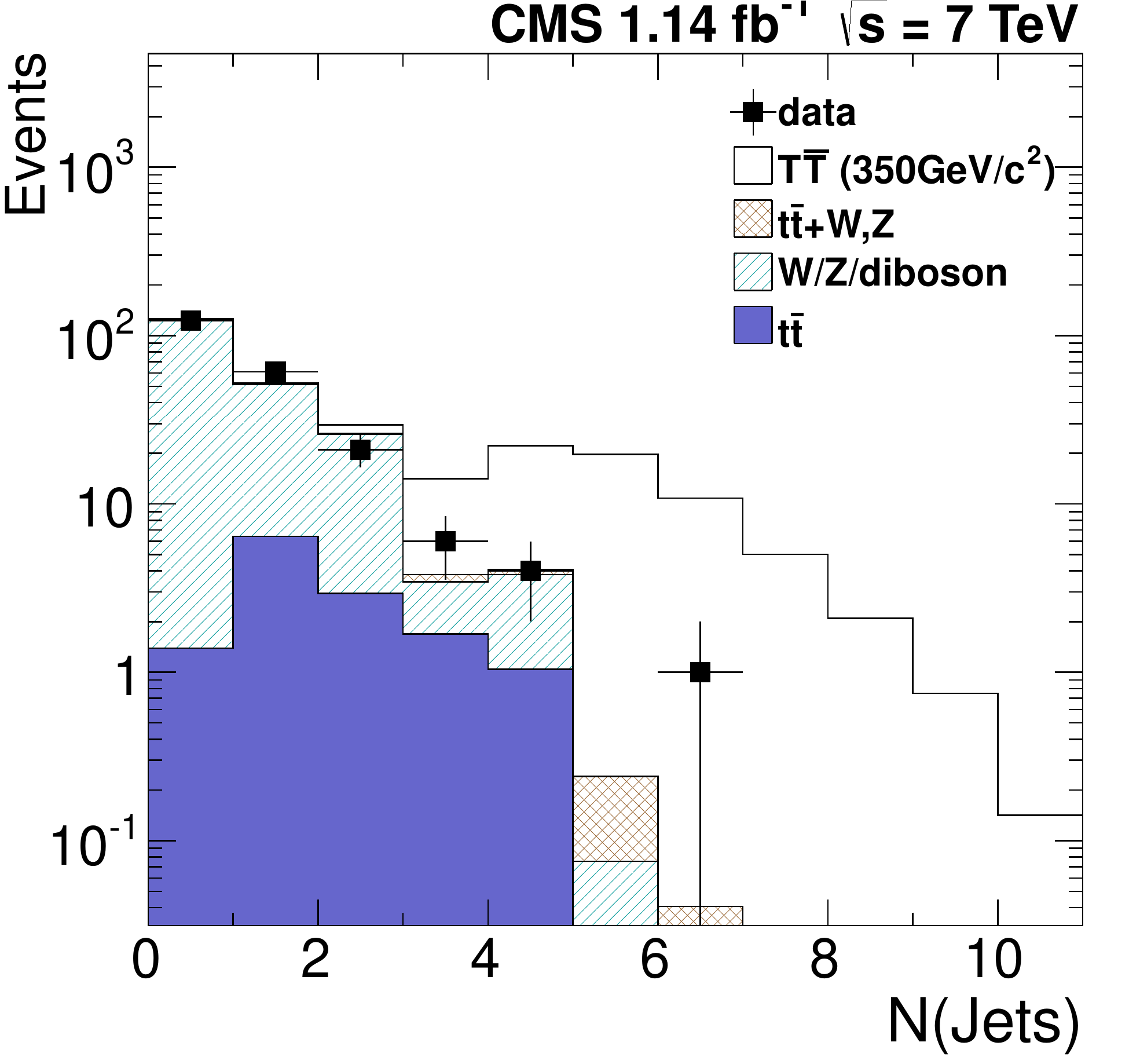}
    \includegraphics[width=0.3\textwidth]{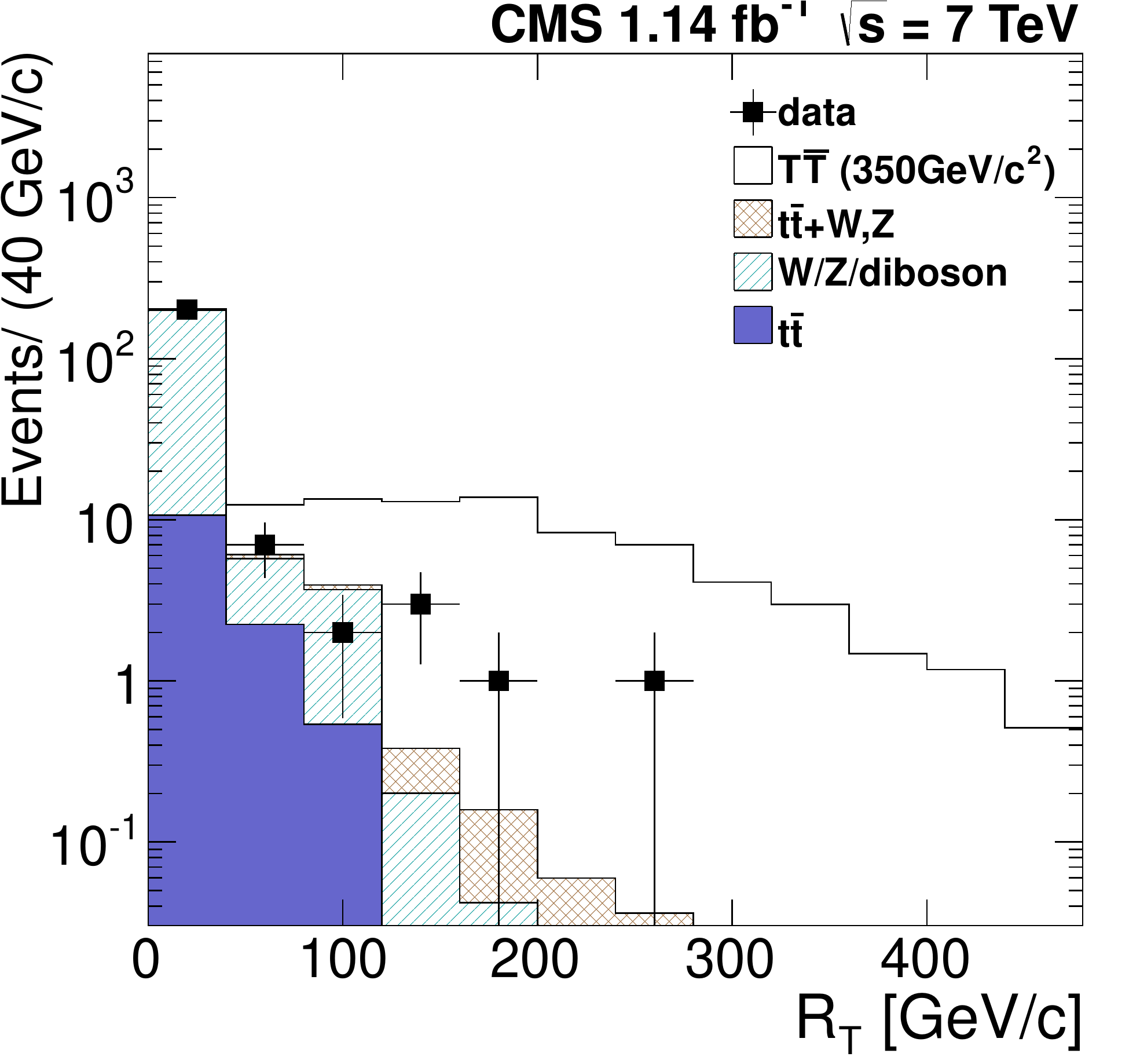}
\caption{The distributions of the invariant mass of two oppositely charged muons or electrons
from data (points) and from Monte Carlo simulations of the backgrounds (colored histograms) and a 350 $\GeVcc$ $\TTbar$
signal (open histograms),
$M_{\ell^+\ell^-}$ (left), jet multiplicity (center), and $R_{\bigT}$ (right) for events
with a reconstructed \cPZ-boson candidate and a charged lepton.
}
\label{fig1}
  \end{center}
\end{figure*}

Simulated event samples are used to estimate the signal efficiencies.
The $\Pp\Pp\to \TTbar X$ process, with up to two additional hard partons, is simulated using the \MADGRAPH~\cite{MADGRAPH} event generator. The result is passed to \PYTHIA (v6.420)~\cite{PYTHIA} for parton showering and hadronization.
Detector simulation is performed using \GEANTfour~\cite{GEANT4}.
The signal efficiencies, excluding the combined branching fractions of 5.4\% from the $\PW$ and $\cPZ$ leptonic decays,
vary from ($14\pm3$)\% to ($36\pm6$)\% as the $\bigT$ mass increases from 250 to 550\GeVcc, where the uncertainties are systematic.
The reduction of signal efficiency for events with a lower \bigT-quark mass is due to the requirement on
$R_{\bigT}$ and the minimum $p_{\bigT}$ threshold for lepton candidates.
Contributions from cascade decays of $\tau$ lepton's are negligible.
The distributions of the dilepton invariant mass, jet multiplicity, and $R_{\bigT}$ for
events with a \cPZ-boson candidate and a charged lepton are displayed in Fig.~\ref{fig1}.
The expected distributions of a \bigT signal with 350\GeVcc mass also shown in Fig.~\ref{fig1} are normalized using the
$\TTbar$ cross section calculated to approximately next-to-next-to-leading order (NNLO) in $\alpha_s$~\cite{Aliev:2010zk}.

After the full selection criteria are applied, two types of background sources
remain in the signal sample:
a) events with two prompt leptons ($B_{2\ell}$) and a non-prompt lepton from a jet; b)
events with three prompt leptons ($B_{3\ell}$).
To estimate the yield of the $B_{2\ell}$ background in data, a method using a
sample of leptons passing looser selection criteria than those described above is introduced.
This type of background is primarily from $\cPZ$ and $\cPqt\cPaqt$ processes.
Electrons chosen with the full selection criteria defined  above are
called ``tight'' electrons. Electron candidates that are above the
same $\pt$ threshold, isolated from other activities in the detector,
but fail the full selection criteria are called ``loose'' electrons.
Similarly, muons chosen with the full selection criteria are ``tight'' muons, while
muon candidates passing the selection criteria defined above except the
requirement on the sum of the $\pt$ of tracks surrounding the muon candidate are
``loose'' muons.
An additional control sample is defined with selection criteria similar to those of the
signal sample, except that the third lepton must only satisfy
the loose lepton requirements.
$\cPZ$ and $\cPqt\cPaqt$ production are the dominant processes also in the control sample, similarly to the signal sample.
The background is estimated using the event yield observed in the control sample,
multiplied by
the probability of a loose lepton in background events passing the tight criteria.
This probability is determined from data by taking
the number of events in a multijet dominant control sample, and dividing the number of events with one loose and one tight lepton by the number of events with two loose leptons.
For electrons this probability is
($2.00\pm0.02$)\% and for muons it is ($18.7\pm0.1$)\%, where the uncertainties are statistical only.
The background yield in the signal sample is estimated to be $3.0\pm0.8$~events, and
the possible presence of signal events in the control sample has a negligible effect.
The small contribution from QCD multijet processes is included in
this estimation.
The method described above  predicts a background contribution in the signal sample that is
consistent with the expectation from simulated standard model event samples.

The contribution of $B_{3\ell}$ background from processes
such as $\cPqt\cPaqt+\cPZ$ and diboson
production is evaluated from simulations using the \MADGRAPH and \PYTHIA generators.
These background processes are irreducible and their contribution amount to
$1.6\pm0.5$ events.
As summarized in Table~\ref{tab:back_estimate},
the total estimated background yield in the signal sample is $4.6\pm1.0$~events, including the systematic uncertainties
described below.
Seven events are observed in data, compatible with the SM expectation.

\begin{table}[t]
\caption{
Predicted number of background events having two prompt leptons ($B_{2\ell}$), estimated using data,
three prompt leptons ($B_{3\ell}$), estimated using simulations, and their sum ($B_\text{total}$)
in each of the trilepton channels, as well as the observed yield in data after applying the full selection
criteria.
The uncertainties shown include both statistical and systematic.
}
\label{tab:back_estimate}
\begin{center}
\begin{tabular}{cccccc}
\hline
\hline
Channel  		& \Pe\Pe\Pe         & \Pe\Pe$\Pgm$     & $\Pgm\Pgm$\Pe   & $\Pgm\Pgm\Pgm$ & Total \\
\hline
$B_{2\ell}$ 	& $0.2^{+0.3}_{-0.2}$ & $0.8\pm0.5$ & $0.9\pm0.4$ & $1.1\pm0.5$ & $3.0\pm0.8$ \\
$B_{3\ell}$ 	& $0.3\pm0.1$ & $0.3\pm0.1$ & $0.5\pm0.2$ & $0.5\pm0.2$ & $1.6\pm0.5$ \\
\hline
$B_\text{total}$	& $0.5\pm0.3$ & $1.1\pm0.5$ & $1.4\pm0.5$ & $1.7\pm0.6$ & $4.6\pm1.0$ \\
\hline
Data 			& 0           & 2           & 2           & 3           & 7 \\
\hline
\hline
\end{tabular}
\end{center} \end{table}

The systematic uncertainties on the signal efficiencies and the
background estimation are summarized in Table~\ref{tab:systematic}.
The uncertainty on the integrated luminosity is estimated to be 4.5\%~\cite{LUMI},
and is included in the limit calculations.
An uncertainty of 2.1\% in the trigger efficiency for
signal events is obtained by comparing the trigger efficiency measured
from data with that measured from the simulated signal sample.
The lepton selection efficiencies computed from $\TTbar$ simulated events
are checked in data using $\cPZ$ samples. The difference between the
efficiencies measured in simulated $\cPZ$ boson and $\TTbar$ signal samples is
taken into account. The resulting uncertainties on the lepton selection
efficiencies are 5.7\% and 7.1\% for electrons and muons, respectively,
giving a total uncertainty of 17\% on the signal selection efficiency, and
an uncertainty of $\pm0.3$ events on the background estimation.
The effects of multiple $\Pp\Pp$ collisions per beam crossing (pileup) are tested with simulations.
Weights are assigned to the simulated events so that the distribution of the number of pileup events matches
the target distribution in data. The associated uncertainty is estimated by varying the weights for different
distributions.
The uncertainty on the parton distribution function (PDF) from CTEQ6~\cite{PDF} and the jet energy
scale~\cite{Chatrchyan:2011ds} and resolution are also
accounted for.
The uncertainty on the background estimation due to the statistical
size of the control samples is $\pm0.7$ events.
The effect of uncertainties on the background cross sections are
considered by varying the normalization of the relevant
processes as follows: $\pm11$\% for $\ttbar
$~\cite{XSEC_tt}, $\pm3$\% ($\pm4$\%) for $\PW$ ($\cPZ$)~\cite{XSEC_WZ}, conservatively $\pm$(27 to 42)\% for dibosons~\cite{MCFM}, and $\pm50$\% for
$\ttbar+\PW/\cPZ$.

\begin{table}[t]
\caption{A summary of relative systematic uncertainties on the signal efficiencies ($\Delta\epsilon/\epsilon$) in percent and estimated background yield.
The uncertainties on the signal efficiency vary with the \bigT-quark mass and these variations are shown by the ranges given in the table.
The uncertainties on the number of background with two prompt leptons~($\Delta B_{2\ell}$), three prompt leptons~($\Delta B_{3\ell}$), and their sum~($\Delta B_\text{total}$) are also summarized.
In all cases, the uncertainties from different sources are summed in quadrature to obtain the total uncertainty, while the correlations
between different background sources are taken into account.
}
\label{tab:systematic}
\begin{center}
\begin{tabular}{lcccc}
\hline
\hline
Source      					& Signal & \multicolumn{3}{c}{Background} \\
                             	& $\Delta\epsilon/\epsilon$[\%]& $\Delta B_{2\ell}$ & $\Delta B_{3\ell}$ & $\Delta B_\text{total}$ \\
\hline
Luminosity         				& 4.5 		& - 	& 0.1 	& 0.1 \\
Trigger efficiency    			& 2.1 		& - 	& - 	& -	  \\
Lepton selection				& 17    	&$<0.01$& 0.3 	& 0.3 \\
Pileup 	               			& 2.3     	& 0.3 	& 0.06 	& 0.4 \\
PDF                    			& 0.2-1.4	& - 	& 0.03 	& 0.03 \\
Jet energy scale/resolution		& 0.8-5.4 	& 0.3 	& 0.2 	& 0.4 \\
Simulated sample statistics  	& 3.0-4.8	& -   	& 0.1 	& 0.1 \\
Control region statistics  		& -			& 0.7 	& - 	& 0.7 \\
Background normalization 		& - 		& 0.2 	& 0.4 	& 0.4 \\
\hline
Total                          	& 18-20     & 0.8 & 0.5 & 1.0 \\
\hline
\hline

\end{tabular}
\end{center}
\end{table}

\begin{table*}[t]
\caption{Summary of the predicted $\TTbar$ cross sections, selection efficiencies, and expected yields for various \bigT masses, normalized to an integrated luminosity of $1.14\fbinv$,
and the observed upper limits at the 95\% confidence level on the cross section.
The expected yields include the combined branching fraction of 5.4\% from the $\PW$ and $\cPZ$ leptonic decays. }
\label{tab:signal}
\begin{center}
\setlength{\tabcolsep}{2.5pt}
\begin{tabular}{lccccccc}
\hline
\hline
$M(\bigT)$ [GeV/$c^2$]	& 250 & 300 & 350 & 400 & 450 & 500 & 550 \\
\hline
Cross-section [pb] 		& 22.6 & 7.99 & 3.20 & 1.41 & 0.662 & 0.330 & 0.171 \\
Efficiency [\%] 		&  $14.4\pm2.8$ & $24.0\pm4.4$ & $29.4\pm5.3$ & $32.8\pm5.8$ & $34.3\pm6.1$ & $32.7\pm5.8$ & $35.6\pm6.3$ \\
Expected yield 			& 200 & 118 & 57.8 & 28.3 & 13.9 & 6.6 & 3.7 \\
\hline
Observed limit [pb] 	& 1.09 & 0.65 & 0.53 & 0.48 & 0.45  & 0.48  & 0.44 \\
\hline
\hline
\end{tabular}
\end{center} \end{table*}

\begin{figure}[t]
  \begin{center}
    \includegraphics[width=\cmsFigWidth]{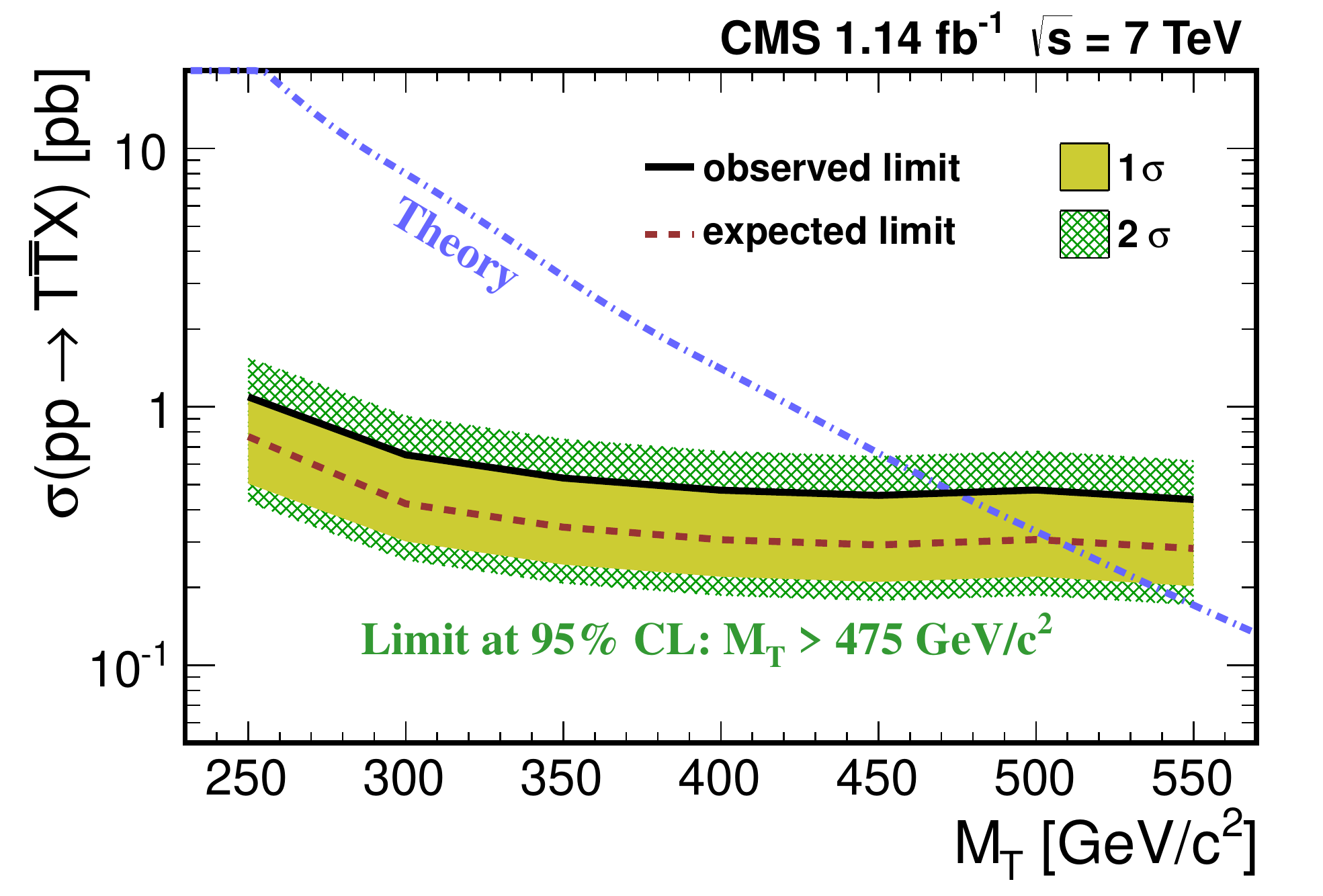}
\caption{
The 95\% confidence level (CL) upper limit on the cross section of the $\Pp\Pp \to\TTbar X$ process, as a function of the \bigT-quark mass.
The branching fraction of $\bigT\to \cPqt\cPZ$ is assumed to be 100\%.
The solid line shows the observed limit.
The dotted line corresponds to the expected limit under a background-only hypothesis.
The solid (hatched) area shows the $\pm1$ ($\pm2$) standard deviation uncertainties on the expected limit.
The dot-dash line shows the value of the theoretical cross section~\cite{Aliev:2010zk} for the $\TTbar$ process.
}
\label{fig:result_exclusion_limit}
  \end{center}
\end{figure}

For each $\bigT$~mass hypothesis from 250 to 550 $\GeVcc$
we present the predicted cross section, selection efficiency, and yield in
Table~\ref{tab:signal}.
Upper limits on the cross section are calculated using a Bayesian method~\cite{BAYESIAN} with a flat prior for the signal
cross section, and a log-normal model for integration
over the nuisance parameters.
The observed upper limit at the 95\% confidence level (CL) on the $\TTbar$ cross section
as a function of the \bigT-quark mass hypotheses
is shown as a solid line in Fig.~\ref{fig:result_exclusion_limit}.
The dotted line gives the expected upper limit on the cross section under a background-only hypothesis, and the solid and hatched areas around it show the $\pm1$ and $\pm2$ standard deviation uncertainties on the expected limit. These were found by producing a large sample of pseudo-experiments in which the expected number of background events was allowed to vary according to its statistical and systematic uncertainties, and the resulting upper limit was then determined.
By comparing the observed $\TTbar$ upper limit with the approximate NNLO calculation of
the $\Pp\Pp \to \TTbar X$ production cross section~\cite{Aliev:2010zk} and assuming a $100\%$ branching fraction
for $\bigT\to \cPqt\cPZ$ decays, a lower limit on the \bigT-quark mass of $475\GeVcc$ is derived at the 95\% confidence level.

In conclusion, using a data sample corresponding to an integrated luminosity of $1.14\fbinv$ collected by the CMS
experiment,
we have searched for a vector-like charge-2/3 \bigT quark that is
pair-produced in $\Pp\Pp$ collisions at a center-of-mass energy
of $7\TeV$ and decays to a top~quark and a \cPZ~boson.
Seven events are observed in data, consistent with $4.6\pm1.0$ events expected from SM processes.
Assuming a $100\%$ branching fraction
for the decay $\bigT\to \cPqt\cPZ$, we exclude a \bigT~quark with a mass less
than $475\GeVcc$ at the $95\%$ confidence level. This is the first search for a pair-produced \bigT quark at hadron colliders.

We wish to congratulate our colleagues in the CERN accelerator departments for the excellent performance of the LHC machine. We thank the technical and administrative staff at CERN and other CMS institutes, and acknowledge support from: FMSR (Austria); FNRS and FWO (Belgium); CNPq, CAPES, FAPERJ, and FAPESP (Brazil); MES (Bulgaria); CERN; CAS, MoST, and NSFC (China); COLCIENCIAS (Colombia); MSES (Croatia); RPF (Cyprus); Academy of Sciences and NICPB (Estonia); Academy of Finland, ME, and HIP (Finland); CEA and CNRS/ IN2P3 (France); BMBF, DFG, and HGF (Germany); GSRT (Greece); OTKA and NKTH (Hungary); DAE and DST (India); IPM (Iran); SFI (Ireland); INFN (Italy); NRF and WCU (Korea); LAS (Lithuania); CINVESTAV, CONACYT, SEP, and UASLP-FAI (Mexico); PAEC (Pakistan); SCSR (Poland); FCT (Portugal); JINR (Armenia, Belarus, Georgia, Ukraine, Uzbekistan); MST and MAE (Russia); MSTD (Serbia); MICINN and CPAN (Spain); Swiss Funding Agencies (Switzerland); NSC (Taipei); TUBITAK and TAEK (Turkey); STFC (United Kingdom); DOE and NSF (USA).

\bibliography{auto_generated}

\cleardoublepage \appendix\section{The CMS Collaboration \label{app:collab}}\begin{sloppypar}\hyphenpenalty=5000\widowpenalty=500\clubpenalty=5000\textbf{Yerevan Physics Institute,  Yerevan,  Armenia}\\*[0pt]
S.~Chatrchyan, V.~Khachatryan, A.M.~Sirunyan, A.~Tumasyan
\vskip\cmsinstskip
\textbf{Institut f\"{u}r Hochenergiephysik der OeAW,  Wien,  Austria}\\*[0pt]
W.~Adam, T.~Bergauer, M.~Dragicevic, J.~Er\"{o}, C.~Fabjan, M.~Friedl, R.~Fr\"{u}hwirth, V.M.~Ghete, J.~Hammer\cmsAuthorMark{1}, S.~H\"{a}nsel, M.~Hoch, N.~H\"{o}rmann, J.~Hrubec, M.~Jeitler, W.~Kiesenhofer, M.~Krammer, D.~Liko, I.~Mikulec, M.~Pernicka, B.~Rahbaran, H.~Rohringer, R.~Sch\"{o}fbeck, J.~Strauss, A.~Taurok, F.~Teischinger, C.~Trauner, P.~Wagner, W.~Waltenberger, G.~Walzel, E.~Widl, C.-E.~Wulz
% ...
\vskip\cmsinstskip
% ...
1:~~Also at CERN, European Organization for Nuclear Research, Geneva, Switzerland\\
% ...

\end{sloppypar}
\end{document}